\documentclass[twocolumn,showpacs,preprintnumbers,amsmath,eqsecnum,amssymb]{revtex4}
\begin{document}
\preprint{IMAFF-RCA-03-07}
\title{Tensorial perturbations in the bulk of inflating brane worlds}
\author{Pedro F. Gonz\'{a}lez-D\'{\i}az}
\affiliation{Centro de F\'{\i}sica ``Miguel A. Catal\'{a}n'', Instituto de
Matem\'{a}ticas y F\'{\i}sica Fundamental,\\ Consejo Superior de
Investigaciones Cient\'{\i}ficas, Serrano 121, 28006 Madrid (SPAIN).}
\date{\today}
\begin{abstract}
In this paper we consider the stability of some inflating
brane-world models in quantum cosmology. It is shown that whereas
the singular model based on the construction of inflating branes
from Euclidean five-dimensional anti-de Sitter space is unstable
to tensorial cosmological perturbations in the bulk, the
nonsingular model which uses a five-dimensional asymptotically
anti-de Sitter wormhole to construct the inflating branes is
stable to these perturbations.
\end{abstract}


\maketitle

\section{Introduction}

There has recently been a lot of interest in inflating brane
worlds which are obtained by gluing at a given slice two copies of
a truncated either anti-de Sitter- or regular wormhole-spacetime
in five dimensions [1,2]. These brane worlds can in both cases
evolve along the cosmological time following the conventional
pattern of most cosmic branes (i.e. the evolution being described
by a Friedmann equation which shows a early dependence on the
square of the energy density [3]), after undergoing a primordial
period of pure de Sitter inflation. It is then clear that the
four-branes resulting in these models are by themselves stable to
all types of cosmological perturbations taking place {\it in} the
branes [4]. However, no investigation has still be undertaken on
the stability of the bulk under the same type of perturbations in
the above two models. In this paper we shall study in some detail
this important missing topic in the case that the cosmological
perturbations originally considered by Lifshitz and Khalatnikov
[5] are extended to a five-dimensional manifold [6]. The main
result of this study is that whereas the construct obtained from
the five-dimensional anti-de Sitter space is unstable to tensorial
cosmological perturbations, the construct which is obtained from a
regular five-dimensional asymptotically anti-de Sitter wormhole is
stable to these perturbations.

We can outline the paper as follows. In Sec. II we discuss the
instabilities present in the instanton constructed from the
five-dimensional anti-de Sitter space. Sec. III contains a
calculation which leads to the conclusion that whereas the
four-dimensional asymptotically anti-de Sitter shows tensorial
instabilities, the five-dimensional asymptotically anti-de Sitter
wormhole bulk of a four-dimensional brane is stable to tensorial
perturbations. We conclude in Sec. IV.

\section{Instabilities of de Sitter brane worlds}

The scale factor of the five-dimensional Garriga-Sasaki model for
an inflating brane world, which describes a spacetime with
topology R $\times$ ${\rm S}^4$, i.e., denoting the metric on the
unit four-sphere by $d\Omega_4^2$,
\[ds^2=dr^2+a(r)^2 d\Omega_4^2 ,\]
is given by [1]
\begin{equation}
a(r)=\ell\sinh(r/\ell) ,
\end{equation}
where $\ell=(-6/\Lambda_5)^{1/2}$ is the ADS radius and $r$ is the
extra fifth coordinate. It was pointed out in Refs. [1,2] that
this solution may lead to a problem. It is that the non-vanishing
components of the five-dimensional Weyl tensor
$C_{\nu\rho\sigma}^{\mu}$ gives rise to the invariant quantity
$C^2\propto a(r)^{-4}$ which may vanish at $r=0$ if we take for
$a(r)$ the solution (2.1). That shortcoming was circumvented by
using the same spacetime topology as in the Garriga-Sasaki model,
but for a scale factor given by
\begin{equation}
a(r)=\left(\frac{\sqrt{\beta}\cosh(2\sqrt{\Lambda_5}r)
-1}{2\Lambda}\right)^{1/2} ,
\end{equation}
(with $\beta=1+4A^2\Lambda_5$) while still producing the same de
Sitter inflating brane-world model.

In this paper it will be seen that the solution used by
Garriga-Sasaki is unstable in still another respect, that of the
gravitational-wave perturbations on the bulk, and that again using
a solution like (2.2) also solves this new problem for
constructing an inflating de Sitter brane world scenario. This can
be explicitly shown by considering the Lifshitz-Khalatnikov
tensorial cosmological perturbations [5] generalized to a
five-dimensional Friedmann-Robertson-Walker manifold [6] for the
scale factors given by Eqs. (2.1) and (2.2). Expanding in
five-dimensional tensor harmonics and taking for the most general
metric perturbations [6]
\[h_{\alpha\beta}=\lambda(\eta)P_{\alpha\beta}+\mu(\eta)Q_{\alpha\beta}+
\sigma(\eta)S_{\alpha\beta}+\nu(\eta)H_{\alpha\beta} ,\] where
$P_{\alpha\beta}$, $Q_{\alpha\beta}$, $S_{\alpha\beta}$ and
$H_{\alpha\beta}$ are tensor harmonics derived from the scalar,
vector and tensor harmonics defined on the four-sphere [6], and
the coefficients $\lambda(\eta)$, $\mu(\eta)$, $\sigma(\eta)$ and
$\nu(\eta)$ are functions of the conformal extra coordinate
$\eta=\int dr/a(r)$. The perturbations originated from the
tensorial five-dimensional gravitational-wave perturbations are
given by coefficient $\nu(\eta)$ which satisfies a differential
equation which in the Lorentzian manifold reads [6]
\begin{equation}
\nu ''+3\frac{a(\eta)'}{a(\eta)}\nu '+\ell(\ell+3)\nu =0 ,
\end{equation}
with $'=d/d\eta$, where $\eta$ is again the conformal coordinate
associated with the extra fifth dimension $r$, i.e. $\eta=\int
dr/a(r)$, with $a(r)$ given by Eq. (2.2) for the present case.
What we shall investigate in this paper is how tensorial
perturbations on the four-sphere $\Omega_4$, represented by the
coefficient $\nu$, evolve along either the "Lorentzian" or
"Euclidean" conformal fifth-coordinate for a generic metric
\[ds^2=a(\eta)^2\left(\pm d\eta^2+d\Omega_4^2\right) ,\] with the
upper sign standing for the Euclidean manifold and the lower one
for the Lorentzian manifold.

In the Euclidean case that corresponds to the metric used by
Garriga and Sasaki [1], we have
\begin{equation}
\nu ''+3\frac{a(\eta)'}{a(\eta)}\nu '=\ell(\ell+3)\nu ,
\end{equation}
where
\begin{equation}
\eta=\eta_*+\frac{1}{2}+\ln\left(\frac{\cosh(\sqrt{\Lambda_5}r)
-1}{\cosh(\sqrt{\Lambda_5}r) +1}\right) ,
\end{equation}
in which $\eta_*$ is an integration constant. The scale factor
expressed in terms of the conformal radial coordinate $\eta$
becomes then
\begin{equation}
a(\eta)=\frac{1}{\sqrt{\Lambda_5}\sinh(\eta-\eta_*)} .
\end{equation}
The differential equation (2.4)can therefore be written
\begin{equation}
\nu ''-3\coth(\eta-\eta_*)\nu '=\ell(\ell+3)\nu .
\end{equation}
We note that even for the zero-mode $\ell=0$ there is an
instability, as the solution to this equation reads
\begin{equation}
\nu=\nu_0 +\nu '_0\left[\frac{1}{3}\cosh^3 (\eta-\eta_*)
-\cosh(\eta-\eta_*)\right] ,
\end{equation}
in which $\nu_0$ and $\nu '_0$ are integration constants. It
follows that for $\eta\rightarrow\infty$ (i.e. as $r\rightarrow
0$) $\nu$ blows up. In the corresponding Lorentzian case this mode
would oscillate along the timelike extra dimension. The conclusion
is thus obtained that the five-dimensional bulk is unstable for
solution (2.1) or (2.6). The stability of the generalized
$d$-dimensional Garriga-Sasaki scenario can also be analyzed
following a similar treatment. Since the Garriga-Sasaki metric for
an arbitrary number of dimensions $d$ is found to be tractable [7]
and expressible as \[ds^2=a(\eta)^2\left(\pm
d\eta^2+d\Omega_{d-1}^2\right) ,\] with $a(\eta)$ given again by
Eq. (2.6) for a $d$-dimensional cosmological constant $\Lambda_d$,
then the differential equation for tensorial perturbations $\nu$
[6] for the Euclidean case becomes \[\nu ''-(d-2){\rm
cotanh}(\eta-\eta_*)\nu '=\ell(\ell+d-2)\nu .\] For $\ell=0$, we
now obtain
\[\nu '=\nu_0 '\sinh^{d-2}(\eta-\eta_*) ,\] with $\nu_0 '$ an
integration constant. It follows that for $\ell=0$ the coefficient
$\nu$ will be generally given as a polynomials of certain powers,
$p\leq d$, of $\cosh(\eta-\eta_*)$ and $\sinh(\eta-\eta_*)$ and
some of their mutual products for odd $d$, and as a similar (not
identical) polynomials plus a term $\pi(\eta-\eta_*)$ (with $\pi$
a constant) for even $d$. In the Lorentzian framework, the
polynomials are similarly given in terms of circular rather than
hyperbolic functions. We see then that the above conclusion for
$d=5$ can be extended for any arbitrary dimension $d$.

\section{Stability of asymptotically anti-de Sitter wormholes}

We shall show now that the above kind of instability problem is no
longer present in case that we use solution (2.2). When we express
such a solution in terms of the conformal radial extra coordinate
\begin{eqnarray}
&&\beta^{1/4}\eta=\beta^{1/4}\eta_* +\nonumber\\
&&F\left[\arcsin\sqrt{\frac{\sqrt{\beta}
\left(\cosh(2\sqrt{\Lambda_5}r)
-1\right)}{\sqrt{\beta}\cosh(2\sqrt{\Lambda_5}r)
-1}},R(\beta)\right] ,
\end{eqnarray}
with
\[R(\beta)=\sqrt{\frac{\sqrt{\beta}+1}{2\sqrt{\beta}}} ,\]
and where again $\eta_*$ is an integration constant and $F$ is the
elliptic integral of the first class [7], the scale factor can be
written
\begin{equation}
a(\eta)=\gamma {\rm nc}\left(\beta^{1/4}\eta|m\right) ,
\end{equation}
in which we have absorbed the constant $\eta_*$ into $\eta$,
\begin{equation}
\gamma=\frac{\sqrt{\beta}-1}{2\Lambda_5} ,
\end{equation}
and ${\rm nc}(x|m)$ is an elliptic function with parameter $m$
[7]. Then, denoting $x=\beta^{1/4}\eta$, we have for the Euclidean
differential equation for the coefficient $\nu$
\begin{equation}
\nu ''+3\beta^{1/4}\frac{{\rm sn}(x|m){\rm dn}(x|m)}{{\rm
cn}(x|m)}\nu '=\ell(\ell+3)\nu ,
\end{equation}
with ${\rm sn}, {\rm dn}$ and ${\rm cn}$ being elliptic functions
as well [7]. Now, for the zero-mode $\ell=0$ we obtain the
analytical solution
\begin{eqnarray}
&&\nu=\nu_0+\frac{\nu_0 '}{2m\beta^{1/4}}\times \nonumber \\
&&\left\{{\rm sn}(x|m){\rm dn}(x|m)
-\frac{m_1}{\sqrt{m}}\arcsin\left(\sqrt{m}{\rm
sn}(x|m)\right)\right\},
\end{eqnarray}
with $m_1=1-m$. It follows that though $\nu$ oscillates along the
entire direction $\eta$, it does not diverge anywhere on that
direction, so that, contrary to what happens for solution (2.1),
the five-dimensional bulk for solution (2.2) appears to be stable
to these zero-mode tensorial perturbations. It is worth noticing,
moreover, that in the Lorentzian case where the solution is given
in terms of the elliptic function ${\rm cn}=1/{\rm nc}$,
\[a(\eta)=\gamma {\rm cn}\left(\beta^{1/4}\eta|m_1\right) ,\]
the solution for the zero-mode for coefficient $\nu$ is the same
as in Eq. (3.5), but with parameter $m$ replaced for parameter
$m_1$ and vice versa. In this way, though the $\ell=0$
gravitational-wave mode is not damped, neither it increases with
Lorentzian time $\eta$.

For $\ell\neq 0$ the differential equation for $\nu$ can be
re-written as
\begin{equation}
\left({\rm nc}(x|m)\nu'\right)'=\ell(\ell+3){\rm nc}^3(x|m)\nu
\end{equation}
in the Euclidean description, and as
\begin{equation}
\left({\rm cn}(x|m_1)\nu'\right)'=\ell(\ell+3){\rm cn}^3(x|m_1)\nu
\end{equation}
in the Lorentzian description, with $x=\beta^{1/4}\eta$ in both
cases. We have been unable to find an analytical solution in
closed form for these differential equations, so that we will
consider the limiting behaviours of $\nu$ as
$\beta^{1/4}\eta\rightarrow 0,2K,...$ and
$\beta^{1/4}\eta\rightarrow K, 3K,...$, where $K$ is the complete
elliptic integral [8]. In the former case for the Euclidean
solution one can approximate $\nu$ to be given by
\begin{equation}
\nu\simeq\nu_0\exp\left(\sqrt{\ell(\ell+3)}\eta\right) ,
\end{equation}
and in the latter Euclidean case $\nu$ tends to generally finite
constant values. Thus, at least for $\eta$-intervals running up to
finite numbers of complete elliptic integrals, we see that there
is no instability arising from these $\ell\neq 0$ modes. For the
Lorentzian description near $\beta^{1/4}\eta\rightarrow 0,2K,...$,
if we choose for the constant $A^2$ the particular value
$A^2=175/(324\Lambda_5)$, we obtain a solution in terms of the
ultraspherical Gegenbauer polynomials [8]
\begin{equation}
\nu\propto C_{\ell}^{3/2}(\eta),
\end{equation}
which tends either to vanish for odd $\ell$, or to a finite
nonzero constant,
\[\nu=(-1)^{\ell/2}\frac{\Gamma\left(\frac{3}{2}+
\frac{\ell}{2}\right)}{\Gamma\left(\frac{3}{2}\right)\left(\frac{3}{2}\right)!},
\]
for even $\ell$, as $\eta$ goes into these limiting values. For
$\beta^{1/4}\eta\rightarrow K, 3K,...$, the Lorentzian solution
would again tend to a generally finite constant. Therefore,
gravitational-wave perturbations do not induce any instability in
the five-dimensional bulk for our nonsingular solution.

It is rather interesting to realize that the stability to
tensorial perturbations of formally the same solutions for the
scale factor will critically depend on the number of dimensions of
the spacetime we deal with. In fact, there exist four-dimensional
asymptotically anti-de Sitter Euclidean wormholes [9,10] which are
characterized by a Robertson-Walker metric and a scale factor
which is exactly the same as that given by Eq. (2.2) for a
cosmological constant $\Lambda_4\equiv\Lambda$. Such wormholes are
usually obtained as the Euclidean solution that corresponds to the
case of a massless scalar field which is conformally coupled to
Hilbert-Einstein gravity plus a cosmological constant $\Lambda$
[9,10]. However, even for the case $\ell=0$, that solution appears
to be unstable to the same kind of tensorial perturbations to
which solution (2.2) is in fact stable. In this case, the
differential equation is modified to be
\begin{equation}
\sigma ''+2\beta^{1/4}\frac{{\rm sn}(x|m){\rm dn}(x|m)}{{\rm
cn}(x|m)}\sigma '=0 .
\end{equation}
The solution to Eq. (3.10) is:
\begin{equation}
\sigma=\sigma_0+\frac{\sigma_0 '}{\beta^{1/4}m}\left[E\left({\rm
am}(\beta^{1/4}\eta),m\right)-m_1\beta^{1/4}\eta\right] ,
\end{equation}
where $E$ is the elliptic integral of the second kind [8], and
${\rm am}$ is the amplitude of the corresponding elliptic function
[8] . We can readily check that in fact this solution diverges as
$\eta\rightarrow\infty$. The counterpart for the Lorentzian baby
universe will nevertheless be stable to gravitational-wave
perturbations and corresponds to the solution
\begin{equation}
\sigma=\sigma_0+\frac{\sigma_0 '}{\beta^{1/4}m_1}\left[E\left({\rm
am}(\beta^{1/4}\eta_L),m_1\right)-m\beta^{1/4}\eta_L\right] ,
\end{equation}
with $\eta_L$ a compact time coordinate.

\section{Conclusions}

The conclusions of this paper are that whereas the Garriga-Sasaki
instanton based on a trivial extension to five or arbitrarily
higher dimensions from the usual four-dimensional anti-de Sitter
instanton and the kind of four-dimensional Euclidean
asymptotically anti-de Sitter wormhole we have just discussed are
unstable to the tensorial gravitational-wave like perturbations,
the baby universe associated with that wormhole and the
nonsingular five-dimensional instanton which was used in [2] to
replace the Garriga-Sasaki solution [1] are both stable to the
Lorentzian counterpart of such perturbations. A caveat to these
conclusions should be mentioned. It is that, since the tensorial
instability of the five-dimensional anti-de Sitter instanton takes
place as $r\rightarrow 0$, this instability would coincide with
the other kind of instability identified for that solution which
arises from the invariant Weyl tensor $C^2$ also at $r=0$. Whether
or not these two apparently distinct kinds of instabilities share
a common ultimate origin is to be investigated. We note, moreover,
that the presence of such instabilities appears to be independent
of the value of $r$ at which the brane is placed. On the other
hand, the five-dimensional wormhole instanton, and hence the brane
constructed from it, are stable as one approaches the limiting
values of $r$ at both $r=0$ and $r=\infty$.

\acknowledgements

\noindent The author thanks Mariam Bouhmadi and Carmen L. Sig\"{u}enza
for useful discussions. This work was supported by DGICYT under
Research Project BMF2002-03758.


\begin{references}
\bibitem {1} J. Garriga and M. Sasaki, Phys. Rev. D62, 043523
(2000); K. Koyama and J. Soda, Phys. Lett. B483, 432 (2000); L.
Anchordoqui and K. Olsen, Mod. Phys. Lett. A16, 1157 (2001); L.
Anchordoqui, C. Nu\~{n}ez and K. Olsen, JHEP 0010, 050 (2000); H.A.
Chamblin and A.S. Reall, Nucl. Phys. B562, 133 (1999); S. Kanno,
M. Sasaki and J. Soda, Prog. Theor. Phys. 109, 357 (2003); R.
Branderberger, G. Geshnizjani and S. Watson, Phys. Rev. D67,
123510 (2003) .
\bibitem {2} M. Bouhmadi and P.F. Gonz\'{a}lez-D\'{\i}az, Phys. Rev. D65,
063510 (2002); M. Bouhmadi-L\'{o}pez, P.F. Gonz\'{a}lez-D\'{\i}az and A. Zhuk,
Class. Quant. Grav. 19, 4863 (2002); P.F. Gonz\'{a}lez-D\'{\i}az, Nucl.
Phys. B619, 646 (2001).
\bibitem {3} P. Bin\'{e}truy, C. Deffayet and D. Langlois, Nucl. Phys.
B565, 269 (2000); P. Bin\'{e}truy, C. Deffayet, U. Ellwanger and D.
Langlois, Phys. Lett. B477, 285 (2000); P.F. Gonz\'{a}lez-D\'{\i}az, Phys.
Lett. B481, 353 (2000).
\bibitem {4} P. Ginsparg and M.J. Perry, Nucl. Phys. B222, 245
(1983).
\bibitem {5} E.M. Lifshitz, J. Phys. 10, 116 (1946); E.M. Lifshitz
and I.M. Khalatnikov, Adv. Phys. 12, 185 (1963).
\bibitem {6} P.F. Gonz\'{a}lez-D\'{\i}az, Phys. Rev. D36, 3651 (1987).
\bibitem {7} P.F. Gonz\'{a}lez-D\'{\i}az, Phys. Lett. B512, 127 (2001).
\bibitem {8} M. Abramowitz and I. Stegun, {\it Handbook of
Mathematical Functions} (Dover, New York, USA, 1964).
\bibitem {9} P.F. Gonz\'{a}lez-D\'{\i}az, {\it Elliptic and Circular
Wormholes}, gr-qc/9306031 .
\bibitem {10} C. Barcel\'{o}, L.J. Garay, P.F.
Gonz\'{a}lez-D\'{\i}az and G.A. Mena Marug\'{a}n, Phys. Rev. D53, 3162 (1996).

\end{references}
\end{document}